# Fine tuning and MOND in a metamaterial "multiverse"

Igor I. Smolyaninov [1], Vera N. Smolyaninova [2]

[1] *Department of Electrical and Computer Engineering, University of Maryland, College Park, MD 20742, USA*

[2] *Department of Physics Astronomy and Geosciences, Towson University, 8000 York Rd., Towson, MD 21252, USA*

**We consider the recently suggested model of a multiverse based on a ferrofluid. When the ferrofluid is subjected to a modest external magnetic field, the nanoparticles inside the ferrofluid form small hyperbolic metamaterial domains, which from the electromagnetic standpoint behave as individual "Minkowski universes" exhibiting different "laws of physics", such as different strength of effective gravity, different versions of MOND and different radiation lifetimes. When the ferrofluid "multiverse" is populated with atomic or molecular species, and these species are excited using an external laser source, the radiation lifetimes of atoms and molecules in these "universes" depend strongly on the individual physical properties of each "universe" via the Purcell effect. Some "universes" are better fine-tuned than others to sustain the excited states of these species. Thus, the ferrofluid-based metamaterial "multiverse" may be used to study models of MOND and to illustrate the fine-tuning mechanism in cosmology.**

## 1. Introduction

Many physical properties of our universe, such as the relative strength of the fundamental interactions, the value of the cosmological constant, etc., appear to be fine-tuned for existence of human life. One possible explanation of this fine tuning assumes existence of a multiverse, which consists of a very large number of individual universes having different physical properties. Intelligent observers populate only a small subset of these universes, which are fine-tuned for life.

While this point of view may not be falsifiable based on astrophysical observations, one possible way to ascertain its viability may rely on condensed matter physics. In particular, it was suggested very recently [1,2] that magnetic nanoparticle-based ferrofluids share some common features with the class of cosmological models based on loop quantum gravity. This analogy relies on the fact that a modest external magnetic field aligns most of the individual magnetic nanoparticles in the ferrofluid into long parallel chains (see Fig.1), so that the ferrofluid becomes a self-assembled hyperbolic metamaterial [3] (an extremely anisotropic uniaxial material, which behaves like a metal in one direction and like a dielectric in the orthogonal direction). Photons in a hyperbolic metamaterial exhibit gravity-like nonlinear optical interactions [4], which may be traced to perturbations of an emergent effective Minkowski spacetime [5] that describes light propagation in such metamaterials. It appears that both loop quantum gravity models and hyperbolic metamaterials may exhibit metric signature phase transitions [6], during which the spacetime metric used to describe the system changes its signature. Moreover, the metric signature transition in a ferrofluid leads to separation of the effective spacetime into a multitude of intermingled Minkowski and Euclidean domains, giving rise to a picture of "metamaterial multiverse" [7]. Inflation-like behaviour appears to be generic within individual Minkowski domains [8]. Thus, ferrofluid-based self-assembled metamaterial geometry captures many features of several cosmological models of the multiverse, such as metric signature transition scenario in loop quantum cosmology [1,2], natural emergence of a large number of Minkowski universes, and inflation. Moreover, in this paper we will demonstrate that individual Minkowski domains in the ferrofluid exhibit different "laws of physics", such as different versions of modified Newtonian dynamics (MOND), different strength of effective gravity, and different radiation lifetimes due to variations in the local Purcell factor, so that ferrofluids may be used to illustrate the fine-tuning mechanism in cosmology. All these effects may be studied via direct microscopic observations.

## 2. Metamaterial "multiverse"

The metric signature transition in ferrofluids (shown in Fig.1) may be understood based on the theory of superparamagnets [9] and the Maxwell-Garnett approximation, which describes electromagnetic properties of wire array metamaterials [10]. Our experiments use the cobalt magnetic fluid 27-0001 from Strem Chemicals, which is composed of 10 nm cobalt nanoparticles in kerosene coated with sodium dioctylsulfosuccinate and a monolayer of LP4 fatty acid condensation polymer. The volume fraction of cobalt nanoparticles in this ferrofluid is $\alpha = 8.2\%$. Magnetic interaction of these nanoparticles is rather weak, so in the absence of external magnetic field the nanoparticles are randomly distributed within the fluid, as illustrated in Fig.1(a). On the other hand, application of a modest external magnetic field leads to formation of nanocolumns aligned along the external field [11], as illustrated in Fig.1(b). In addition, depending on the magnetic field magnitude, solvent used, and the nanoparticle concentration, the ferrofluid undergoes phase separation into nanoparticle rich and nanoparticle poor phases [3], which is easy to observe in microscopic images of the ferrofluid, as illustrated in Fig.2(b): the typical spatial scale of phase separation is 0.1-1 $\mu$m.

Ferrofluids subjected to external magnetic field are known to exhibit classical superparamagnetic behaviour, which is a form of magnetism that is exhibited by magnetic materials consisting of small ferromagnetic or ferrimagnetic nanoparticles. When an external magnetic field $H$ is applied to an assembly of magnetic nanoparticles, their magnetic moments tend to align along the applied field, leading to a net magnetization [9]:

$$M(H,T) = \alpha\mu\tanh\left(\frac{\mu H}{kT}\right) , \qquad (1)$$

where $\alpha$ is the nanoparticle concentration, $\mu$ is their magnetic moment, and $T$ is the temperature. Ferrofluids are known from the experiment to obey the same relationship. However, as illustrated in Fig.1(b) at small magnetic fields not all the nanoparticles in the ferrofluid are aligned into chains, and these free floating nanoparticles are much less aligned by the weak external field due to absence of the neighbors. Therefore, it is reasonable to assume that the volume fraction of the nanoparticles aligned into chains $\alpha(H,T)$ is

$$\alpha(H,T) = \alpha_\infty \tanh\left(\frac{\mu H}{kT}\right) \qquad (2)$$

where the limiting volume fraction $\alpha_\infty = 0.082$ is reached at a very large magnetic field when all the available nanoparticles in the ferrofluid participate in the chain formation.

Let us now consider the electromagnetic properties of the ferrofluid within the scope of this model. They may be understood based on the Maxwell-Garnett approximation via the dielectric permittivities $\varepsilon_m$ and $\varepsilon_d$ of cobalt and kerosene, respectively. At small magnetic fields the difference $\alpha_\infty - \alpha(H.T)$ describes cobalt nanoparticles, which are not aligned and distributed homogeneously inside the ferrofluid. Dielectric polarizability of these nanoparticles may be included into $\varepsilon_d$, leading to slight increase in its value. When the nanoparticles are aligned into chains, the diagonal components of the ferrofluid permittivity may be calculated using the Maxwell-Garnett approximation as follows [10]:

$$\varepsilon_z = \varepsilon_2 = \alpha(H,T)\varepsilon_m + (1-\alpha(H,T))\varepsilon_d \qquad (3)$$

$$\varepsilon_x = \varepsilon_y = \varepsilon_1 = \frac{2\alpha(H,T)\varepsilon_m\varepsilon_d + (1-\alpha(H,T))\varepsilon_d(\varepsilon_d+\varepsilon_m)}{(1-\alpha(H,T))(\varepsilon_d+\varepsilon_m) + 2\alpha(H,T)\varepsilon_d} \qquad (4)$$

where z direction is aligned with the direction of external magnetic field. As far as the magnetic permeability is concerned, at the visible and infrared frequencies the ferrofluid may be considered as a non-magnetic (µ=1) medium. Calculated wavelength dependencies of $\varepsilon_2$ and $\varepsilon_1$ at $\alpha(H,T) = \alpha_\infty = 8.2\%$ are plotted in Fig. 2(a). These calculations are based on the optical properties of cobalt in the infrared range [12]. While $\varepsilon_1$ stays positive and almost constant, $\varepsilon_2$ changes sign to negative around $\lambda =$ 1µm. So in a very strong magnetic field the ferrofluid becomes a hyperbolic metamaterial at $\lambda > 1$µm.

On the other hand, Eq (3) demonstrates that if the volume fraction of cobalt nanoparticles varies depending on either magnetic field or temperature, the change of sign of $\varepsilon_2$ occurs at some critical value $\alpha_H$:

$$\alpha(H,T) > \alpha_H = \frac{\varepsilon_d}{\varepsilon_d - \varepsilon_m}, \qquad (5)$$

so that the ferrofluid becomes a hyperbolic metamaterial [13,14] above the critical volume fraction. Alignment of cobalt nanoparticle chains along the direction of external magnetic field is indeed clearly revealed by microscopic images of the ferrofluid, as shown in Fig.2(b,c).

Propagation of monochromatic extraordinary photons inside the ferrofluid is described by the wave equation, which is formally equivalent to a 3D Klein-Gordon equation for a massive field $\varphi_\omega = E_z$ in a 3D Minkowski spacetime:

$$-\frac{\partial^2 \varphi_\omega}{\varepsilon_1 \partial z^2} + \frac{1}{(-\varepsilon_2)}\left(\frac{\partial^2 \varphi_\omega}{\partial x^2} + \frac{\partial^2 \varphi_\omega}{\partial y^2}\right) = \frac{\omega_0^2}{c^2}\varphi_\omega \tag{6}$$

in which the spatial coordinate $z$ behaves as a timelike variable, and $E_z$ is the z-component of electric field in the wave [5]. Eq.(6) exhibits effective Lorentz invariance under the coordinate transformation

$$z' = \frac{1}{\sqrt{1 - \frac{\varepsilon_1}{(-\varepsilon_2)}\beta}}(z - \beta x) \tag{7}$$

$$x' = \frac{1}{\sqrt{1 - \frac{\varepsilon_1}{(-\varepsilon_2)}\beta}}\left(x - \beta \frac{\varepsilon_1}{(-\varepsilon_2)}z\right),$$

where $\beta$ is the effective boost. Similar to our own Minkowski spacetime, the effective Lorentz transformations in the $xz$ and $yz$ planes form the Poincare group together with translations along $x$, $y$, and $z$ axis, and rotations in the $xy$ plane. Thus, the wave equation (6) describes world lines of massive particles which propagate in an effective 2+1 dimensional Minkowski spacetime. Components of the metamaterial dielectric tensor define the effective metric coefficients $g_{ik}$ of this spacetime: $g_{00} = -\varepsilon_1$ and $g_{11} = g_{22} = -\varepsilon_2$. Nonlinear optical Kerr effect bends this spacetime resulting in the effective gravitational interaction between extraordinary photons [4]. Thus, transition to a hyperbolic state in the ferrofluid, which is described by Eq.(5), may be understood as transition of the effective "electromagnetic spacetime" from Euclidean to Minkowski state. This transition may be driven by either temperature or external magnetic field. Since both parameters may experience variations across the volume of the ferrofluid, it is natural to expect that near the transition the ferrofluid separates into large number of individual Minkowski and Euclidean domains. Indeed, such domains may be clearly revealed in Fig.2(c), which is obtained from the experimental image of light intensity $I(\vec{r})$ in Fig.2(b) by numeric differentiation $|I(\vec{r})-I(\vec{r}+\vec{d}/2)|$, where $\vec{d}$ is the periodicity of stripes visible in Fig.2(b). This numeric procedure reveals the stripe contrast, and emphasizes ferrofluid regions having larger volume fractions of nanoparticles $\alpha(H,T)$ aligned into the nanochains. According to Eq.(5), these regions behave as "Minkowski domains" separated by regions of Euclidean spacetime, leading to a picture of "metamaterial multiverse".

Here it is interesting to note that the effective metric near the Minkowski domain walls emulates cosmological inflation. Indeed, based on Eq.(6) it is clear that the factor $(-\varepsilon_2)$ plays the role of a scale factor of the effective Minkowski spacetime

$$ds^2 = -\varepsilon_1 dz^2 + (-\varepsilon_2)(dx^2 + dy^2) \qquad (8)$$

(since according to eq.(4) $\varepsilon_1$ is positive and almost constant, as plotted in Fig.2(a)). In the case of $z$-dependent $\varepsilon_1=\varepsilon_x=\varepsilon_y$ and $\varepsilon_2=\varepsilon_z$ the electromagnetic field separation into the ordinary and the extraordinary components remains well defined [15,16]. Taking into account $z$ derivatives of $\varepsilon_1$ and $\varepsilon_2$, eq.(6) becomes

$$-\frac{\partial^2 \varphi_\omega}{\varepsilon_1 \partial z^2} + \frac{1}{(-\varepsilon_2)}\left(\frac{\partial^2 \varphi_\omega}{\partial x^2} + \frac{\partial^2 \varphi_\omega}{\partial y^2}\right) + \left(\frac{1}{\varepsilon_1^2}\left(\frac{\partial \varepsilon_1}{\partial z}\right) - \frac{2}{\varepsilon_1 \varepsilon_2}\left(\frac{\partial \varepsilon_2}{\partial z}\right)\right)\left(\frac{\partial \varphi_\omega}{\partial z}\right) +$$
$$+\frac{\varphi_\omega}{\varepsilon_1 \varepsilon_2}\left(\frac{1}{\varepsilon_1}\left(\frac{\partial \varepsilon_1}{\partial z}\right)\left(\frac{\partial \varepsilon_2}{\partial z}\right) - \left(\frac{\partial^2 \varepsilon_2}{\partial z^2}\right)\right) = \frac{\omega_0^2}{c^2}\varphi_\omega \qquad (9)$$

Since $\varepsilon_1$ is approximately equal to $\varepsilon_d$ (as evident from Fig.2(a)) its derivatives may be neglected. If we also neglect the second derivative of $\varepsilon_2$, the wave equation for the extraordinary field $\varphi_\omega = E_z$ may be re-written as

$$-\frac{\partial^2 \varphi_\omega}{\varepsilon_1 \partial z^2} + \frac{1}{(-\varepsilon_2)}\left(\frac{\partial^2 \varphi_\omega}{\partial x^2} + \frac{\partial^2 \varphi_\omega}{\partial y^2}\right) - \frac{2}{\varepsilon_1 \varepsilon_2}\left(\frac{\partial \varepsilon_2}{\partial z}\right)\left(\frac{\partial \varphi_\omega}{\partial z}\right) = \frac{\omega_0^2}{c^2}\varphi_\omega \qquad (10)$$

It is easy to verify that the latter equation coincides with the Klein-Gordon equation

$$\frac{1}{\sqrt{-g}}\frac{\partial}{\partial x^i}\left(g^{ik}\sqrt{-g}\frac{\partial \psi}{\partial x^k}\right) = \frac{m^2 c^2}{\hbar^2}\psi \qquad (11)$$

for a massive particle in a gravitational field [17] described by the metric coefficients $g_{00} = -\varepsilon_1$ and $g_{11} = g_{22} = -\varepsilon_2$. Indeed, for a wave field $\psi = (-\varepsilon_2)^{1/2}\phi$ Eq.(10) is reproduced if we neglect the second derivative terms in Eq.(11).

Let us assume that the local temperature distribution inside the ferrofluid may be described as

$$T = T_c - z\nabla T = T_c - \eta z \,, \qquad (12)$$

where $T_c$ is the temperature of the metric signature transition (the point where $\varepsilon_2 = 0$). Since z coordinate plays the role of time in the effective spacetime described by eq.(8), the temperature gradient $\eta$ will result in formation of a Minkowski domain, such as the one shown in the experimental image in Fig.3(a). Based on Eq.(3), the experimentally measured image contrast $|I(\vec{r}) - I(\vec{r} + \vec{d}/2)| \sim \alpha(H,T)$ in Fig.2(c) is proportional to the scale factor $(-\varepsilon_2)$ of the effective spacetime within the domain (see Eq.(8)). The cross section of Fig.3(a) along the arrow indicating the effective time direction indeed shows very fast electromagnetic spacetime expansion near the Minkowski domain wall.

The behaviour of the experimentally measured scale factor in Fig.3(b) may be compared to the theoretical prediction based on Eqs.(2-4). We may assume that $-\varepsilon_m \gg \varepsilon_d$, and that $\alpha(H,T)$ is small. In such a case the effective metric coefficients are:

$$g_{11} = g_{22} = -\varepsilon_2 \approx -\alpha_\infty \tanh\left(\frac{\mu H}{kT}\right)\varepsilon_m - \varepsilon_d \tag{13}$$

$$g_{00} = -\varepsilon_1 \approx -\varepsilon_d\left(1 + 2\alpha_\infty \tanh\left(\frac{\mu H}{kT}\right)\right) \tag{14}$$

The effective spacetime appears to be a Minkowski one if the temperature is low enough, or the magnetic field is strong enough, so that

$$\alpha_\infty \tanh\left(\frac{\mu H}{kT}\right) > \frac{\varepsilon_d}{(-\varepsilon_m)} \tag{15}$$

As evident from Fig.3(b), the measured data show good agreement with a theoretical fit obtained based on eq.(13) using the optical properties of cobalt [12] at $\lambda=1.5$ μm. As the ferrofluid temperature falls away from the $T_c$ boundary at z=0, Eq.(13) predicts that the scale factor of the effective Minkowski spacetime increases sharply as a function of z. It is remarkable that this "cosmological" spacetime expansion may be visualized directly using an optical microscope. It is also interesting to note that the physical vacuum appears to exhibit hyperbolic metamaterial properties when subjected to a very strong magnetic field [18,19]. Thus, the ferrofluid in an external magnetic field provides us with an easily accessible experimental system, which may be used to illustrate some properties of vacuum and the hypothesized physics of multiverse. Below we will demonstrate that it also lets us trace the emergence of analogue gravity and the modified Newtonian dynamics (MOND) from the well-understood microscopic degrees of freedom.

## 3. Analogue gravity and MOND in ferrofluids

Since the language of effective spacetime appears to be quite efficient at describing light propagation inside the hyperbolic ferrofluid, it is also natural to express nonlinear optical interaction of photons via the Kerr effect, as "bending" of this spacetime, which results in effective gravitational interaction between extraordinary photons [4]. Indeed, when the nonlinear optical effects become important, the dielectric tensor of the ferrofluid may be written as

$$\varepsilon_{ij} = \chi_{ij}^{(1)} + \chi_{ijlm}^{(3)} E_l E_m + ... \quad , \tag{16}$$

where the second term provides gravity-like coupling between the effective metric and the energy-momentum tensor. It was demonstrated in ref.[4] that in the weak field (Newtonian) limit the effective gravitational constant may be expressed as

$$\gamma = \frac{c^2 \omega^2 \varepsilon_2}{4} \chi^{(3)} \tag{17}$$

Eq.(17) establishes connection between the effective gravitational constant $\gamma$ and the third order nonlinear susceptibility $\chi^{(3)}$ of the ferrofluid. Since $\varepsilon_2 < 0$, the sign of $\chi^{(3)}$ must be negative for the effective gravity to be attractive. This condition is satisfied naturally in most liquids, and in particular, in kerosene. Because of the large and negative thermo-optic coefficient inherent to most liquids, heating produced by partial absorption of the propagating beam translates into a significant decrease of the refractive index at higher light intensity. The thermal origin of this effective gravity looks interesting in light of the modern advances in gravitation theory [20,21], which strongly indicate that the classic general relativity description of gravity results from thermodynamic effects. It is also interesting to note that variations of the dielectric permittivity component $\varepsilon_2$ within the individual Minkowski domains (or "universes") described by Eqs.(3,13) lead to different strength of the effective gravity in different domains.

Let us also analyze variations of the effective modified Newtonian dynamics (MOND) [22], which arise naturally in different Minkowski domains inside the ferrofluid. The effective "second law of Newtonian mechanics" for photons inside a hyperbolic ferrofluid domain may be derived based on the extraordinary photon dispersion law

$$\frac{\omega^2}{c^2} = \frac{k_z^2}{\varepsilon_1} - \frac{k_x^2 + k_y^2}{(-\varepsilon_2)} \quad (18)$$

which follows from Eq.(6), and on the explicit expressions for $\varepsilon_1$ and $\varepsilon_2$ given by Eqs.(13,14). These equations result in the following expression for $k_z^2$:

$$k_z^2 = \frac{\omega^2}{c^2}\varepsilon_d + 2\frac{\omega^2}{c^2}\varepsilon_d \alpha_\infty \tanh\left(\frac{\mu H}{kT}\right) + (k_x^2 + k_y^2)\varepsilon_d \left(\frac{1 + 2\alpha_\infty \tanh\left(\frac{\mu H}{kT}\right)}{-\alpha_\infty \varepsilon_m \tanh\left(\frac{\mu H}{kT}\right) - \varepsilon_d}\right) \quad (19)$$

Note that $E = \hbar c k_z$ plays the role of energy, while $\vec{P}^2 = \hbar^2(k_x^2 + k_y^2)$ plays the role of the momentum squared in an effective 2+1 dimensional Minkowski spacetime introduced in Section 2, while the effective mass of the extraordinary photon appears to be $m = \hbar \omega \varepsilon_d^{1/2}/c^2$. Using these newly introduced variables, Eq.(19) may be re-written as

$$E^2 = m^2 c^4 \left(1 + 2\alpha_\infty \tanh\left(\frac{\mu H}{kT}\right)\right) + \vec{P}^2 c^2 \left(\frac{1 + 2\alpha_\infty \tanh\left(\frac{\mu H}{kT}\right)}{-\alpha_\infty \varepsilon_m \tanh\left(\frac{\mu H}{kT}\right) - \varepsilon_d}\right) \quad (20)$$

Further introducing a "non-relativistic" effective energy $E^* = E - mc^2$, Eq.(20) may be re-written in the "non-relativistic form" as

$$E^* = mc^2 \alpha_\infty \tanh\left(\frac{\mu H}{kT}\right) + \frac{\vec{P}^2}{2m}\left(\frac{1 + 2\alpha_\infty \tanh\left(\frac{\mu H}{kT}\right)}{-\alpha_\infty \varepsilon_m \tanh\left(\frac{\mu H}{kT}\right) - \varepsilon_d}\right) \quad (21)$$

The first term in this equation plays the role of the "gravitational potential energy" $m\phi$, which makes it apparent that the spatial gradients of $T$ and $H$ behave as an effective gravitational field in this model. The second term may be identified as the "kinetic energy" of the photon. The "Newton's second law" may be obtained using the Hamilton's equation as

$$\vec{a} = \frac{d}{dt^*}\left(\frac{dE^*}{d\vec{P}}\right) = \frac{d}{mdt^*}\left(\vec{P}\left(\frac{1 + 2\alpha_\infty \tanh\left(\frac{\mu H}{kT}\right)}{-\alpha_\infty \varepsilon_m \tanh\left(\frac{\mu H}{kT}\right) - \varepsilon_d}\right)\right) \quad (22)$$

where the role of effective time $t^*$ is played by the z coordinate. Since $T$ and $H$ may depend on z within the individual Minkowski domains, the different versions of MOND in these Minkowski domains (or "universes") will be different from each other, and will not coincide with the conventional $F=ma$ form of the second law. Moreover if we re-write Eq.(22) via the effective gravitational potential $\phi$ as

$$\frac{d}{dt^*}\left(\vec{P}\left(\frac{1 + \frac{2\phi}{c^2}}{-\varepsilon_m \frac{\phi}{c^2} - \varepsilon_d}\right)\right) = m\vec{a} \quad (23)$$

and recall that according to the virial theorem the average potential energy of a bound system is proportional to its average kinetic energy, it is clear that the gravitational skipdynamics of bound systems in the ferrofluid would differ considerably from the Newtonian limit if $d\phi/dt^* \sim d(\tanh(\mu H/kT))/dz \neq 0$. The latter condition is always satisfied in the presence of "cosmological expansion" near the Minkowski domain boundaries.

In particular, let us consider the rotation curves of test objects in the effective gravitational field of a central massive body. As has been noted above, in our case the role of the "central mass" *M* and the "test bodies" *m* are played by the light beams, which trajectories behave as world lines in the 2+1 dimensional effective spacetime inside the ferrofluid. In Newtonian 2D gravity the rotation curves are "flat":

$$\frac{mv^2}{r} = \gamma \frac{mM}{r} \qquad (24)$$

so that

$$v = \sqrt{\gamma M} \qquad (25)$$

and does not depend on *r*. Variations of $\phi$ in Eq.(23) as a function of *t\*=z* and *r* near the Minkowski domain boundaries will obviously change these flat rotation curves. However, the 2+1 dimensional character of the effective Minkowski spacetime in the ferrofluid makes the MOND analogies limited, and does not allow realistic modeling and study of such important issues as galaxy rotation curves in the 3+1 dimensional spacetime.

## 4. "Fine tuning" of radiative lifetime in ferrofluids

Let us now consider the difference in radiation lifetimes of excited atoms and molecules inside different Minkowski domains, which is another example of different "laws of physics" which operate in different regions of the ferrofluid-based "metamaterial multiverse". When the ferrofluid "multiverse" is populated with atomic or molecular species, and these species are excited using an external laser source, the radiation lifetimes of atoms and molecules in these "universes" depend strongly on the individual physical properties of each "universe" via the Purcell effect [23,24]. Some "universes" are better fine-tuned than others to sustain the excited states of these species. Thus, the ferrofluid-based metamaterial "multiverse" may be used to illustrate the fine-tuning mechanism in cosmology.

The Purcell effect is the enhancement of a fluorescent molecule's spontaneous emission rate by its environment, which is determined mainly by the enhancement of the photonic density of states. According to the Fermi's golden rule, the transition rate between the excited and the ground state of an atom or molecule is proportional to the density of final states:

$$\Gamma_{i \to f} = \frac{2\pi}{\hbar} \left| \langle f | H_{int} | i \rangle \right|^2 \rho \quad , \tag{26}$$

where $H_{int}$ is the perturbing Hamiltonian, and $\rho$ is the density of final states. As has been shown in [6], the density of photonic states diverges in the hyperbolic metamaterials. This follows from the extraordinary photon dispersion law (Eq.(18)), which describes a hyperboloid in the phase space. Since the absolute value of the photon k-vector is not limited, the phase space volume between two hyperboloids corresponding to different values of frequency is formally infinite in the effective medium approximation. This divergence leads to a very large density of photonic states

$$\Gamma^{meta} \approx \frac{\mu^2 k_{max}^3}{2\hbar} \frac{2\sqrt{-\varepsilon_1 \varepsilon_2}}{(1 - \varepsilon_1 \varepsilon_2)} \tag{27}$$

for every frequency where different components of the dielectric permittivity have opposite signs [23]. Since, the effective medium description eventually fails at the point when the wavelength of the propagating mode becomes comparable to the distance between the nanocolumns $d$ (see Fig.1(b)), this introduces a wave number cut-off at

$$k_{max} = 1/d \qquad (28)$$

of this broadband divergence.

Since its theoretical prediction in [23], the broadband Purcell effect has been indeed observed in multiple experiments (see for example ref. [24]). As demonstrated in Fig.4, narrow absorption lines of the ferrofluid in the long wavelength infrared (LWIR) range also exhibit the broadband Purcell effect. The inset in Fig.4 shows FTIR transmission spectrum of the diluted ($\alpha_\infty$=0.8%) ferrofluid in zero magnetic field. It reveals strong kerosene absorption lines, which separate low absorption regions of the LWIR spectrum where the Purcell effect measurements have been performed. Comparison of the FTIR transmission spectra of the ferrofluid with and without external magnetic field shows clear influence of the Purcell effect on the narrow absorption lines of the ferrofluid, which are marked by green arrows. It is clear from Fig.4 that application of the external magnetic field, which leads to formation of the hyperbolic metamaterial structure, makes the radiation lifetime shorter, leading to observation of broader and less pronounced absorption lines. For example, the broadening of the 8.5 μm absorption line corresponds to ~2 times shorter radiation lifetime. The more pronounced Purcell effect at longer wavelengths may be explained by better plasmonic properties of the cobalt nanoparticles at these wavelengths.

Eqs. (27,28) demonstrate that the Purcell effect in different Minkowski domains inside the ferrofluid is affected by such parameters as local variations of cobalt nanoparticle concentration $\alpha$, external magnetic field $H$ and temperature $T$ (via their influence on $\varepsilon_1$ and $\varepsilon_2$ – see Eqs.(3,4)), as well as local nanocolumn periodicity via the wave vector cutoff. The Minkowski domain size also affects the Purcell effect via the effective cavity dimensions. Considerable variations of the radiation lifetime due to variations in Purcell effect in different Minkowski domains lead to variations of IR absorption, and hence transmission of the domains, which can be easily observed using IR microscopy.

For experimental realization of such microscopic observations we have chosen to operate our setup at 1550 nm using a readily available laser source in transmission mode. A FLIR infrared camera has been used to capture the microscopic images as a function of external magnetic field. This spectral range appears to be quite convenient since it is located above the metric signature transition in cobalt ferrofluid at large magnetic fields (see Fig.2a), it guarantees sufficient spatial resolution for Minkowski domain observation, and because of availability of common C-H, O-H, and N-H absorption bands near 1550 nm. The common C-H absorption bands are located around 1600 nm, N-H bands are located around 1550 nm, and O-H bands are located around 1450 nm, so that IR spectroscopy at these bands is widely used to detect fatty acids [25], which are used as the cobalt nanoparticle coating. As shown in Fig. 5, experimental microscopic images observed at 1550 nm indeed exhibit considerable contrast between different Minkowski domains. For example, cross section of the microscopic image in Fig.5b along the yellow line, which is shown in Fig.5c shows more than two fold variation in Minkowski domain transmission. The observed difference in infrared absorption between different domains indicates at least two fold difference in radiation lifetime between the darker and brighter looking domains. The excited states of the fatty acids appear to have considerably longer lifetime in brighter domains, so that these brighter "universes" appear to be better fine-tuned for their existence. This effect appears to be somewhat analogous to the proposed fine tuning mechanism in multiverse cosmology.

**5. Concluding remarks**

In conclusion, we have demonstrated that ferrofluid-based self-assembled metamaterial geometry captures many features of several cosmological models of the multiverse, such as metric signature transition scenario in loop quantum cosmology [1,2], natural emergence of a large number of Minkowski universes, and inflation. Moreover, we have demonstrated that individual Minkowski domains in the ferrofluid exhibit different "laws of physics", such as different versions of modified Newtonian dynamics (MOND), different strength of effective gravity, and different radiation lifetimes due to variations in the local Purcell effect, so that ferrofluids may be used to illustrate the fine-tuning mechanism in cosmology. All these effects may be studied via direct microscopic observations.

Unfortunately, the described analogy between the extraordinary light propagation inside the ferrofluid and the dynamics of massive particles in Minkowski spacetime is far from being perfect. The main difficulty comes from the cross-coupling between extraordinary and ordinary light inside the ferrofluid, which may be caused by domain interfaces and internal defects. Since ordinary light does not obey the same wave equation (6), such a cross-coupling breaks the effective Lorentz symmetry (7) of the system. In addition, such a model is necessarily limited to 2+1 spacetime dimensions. Nevertheless, despite these limitations the developed condensed matter model of the cosmological multiverse appears to be quite interesting, since it is able to replicate many of its hypothesized features in the laboratory setting.

**Acknowledgments**

This work is supported by the NSF grant DMR-1104676.

**Figure Captions**

**Fig. 1** Schematic geometry of the ferrofluid before (a) and after (b) application of external magnetic field. Before the field is applied, the "optical spacetime" in the ferrofluid has Euclidean character. Alignment of magnetic nanoparticles into chains by the external magnetic field H leads to the metric signature change: the effective spacetime switches from Euclidean to Minkowski.

**Fig. 2** (a) Wavelength dependencies of the real parts of $\varepsilon_z = \varepsilon_2$ and $\varepsilon_{x,y} = \varepsilon_1$ for a cobalt nanoparticle-based ferrofluid at $\alpha = 8.2\%$ volume fraction of nanoparticles. While $\varepsilon_x$ and $\varepsilon_y$ stay positive and almost constant, $\varepsilon_z$ changes sign to negative around $\lambda = 1\mu m$. (b) Microscopic image $I(\vec{r})$ of cobalt nanoparticle-based ferrofluid in external magnetic field reveals nanoparticle alignment along the field direction. (c) Image of the stripe contrast $|I(\vec{r}) - I(\vec{r} + \vec{d}/2)|$, where $\vec{d}$ is the periodicity of stripes visible in (b). This numeric procedure emphasizes ferrofluid regions having larger volume fractions of nanoparticles $\alpha(H,T)$ aligned into the nanochains. According to Eq.(5), these regions behave as "Minkowski domains" separated by regions of Euclidean spacetime.

**Fig. 3** (a) Magnified image of one of the Minkowski domains from Fig.2(c) illustrates inflation-like expansion of the effective spacetime near the domain wall. (b) Measured and theoretically calculated dependencies of the spacetime scale factor $-\varepsilon_2$ on the effective time. The calculations are based on analysis of the measured stripe contrast in (a). The measured data are compared with theoretical calculations performed at $\lambda = 1.5$ $\mu m$ based on eq.(13).

**Fig. 4** Comparison of the FTIR transmission spectra of the ferrofluid with and without external magnetic field shows clear influence of the Purcell effect on the narrow

absorption lines of the ferrofluid. Application of the external magnetic field, which leads to formation of the hyperbolic metamaterial structure, makes the radiation lifetime shorter, leading to broader and less pronounced absorption lines (marked by green arrows). The inset shows FTIR transmission spectrum of the diluted ($\alpha_\infty$=0.8%) ferrofluid in zero magnetic field. It reveals strong kerosene absorption lines, which separate low absorption regions of the LWIR spectrum where the Purcell effect measurements have been performed.

**Fig. 5** (a,b) Microscopic images of the ferrofluid in external magnetic field observed in transmission mode at 1550 nm exhibit considerable contrast between different Minkowski domains. (c) Cross section of microscopic image (b) along the yellow line shows more than two fold variation in Minkowski domain transmission, which indicates more than two fold variation in radiation lifetime of the excited states of fatty acid molecules in these domains.

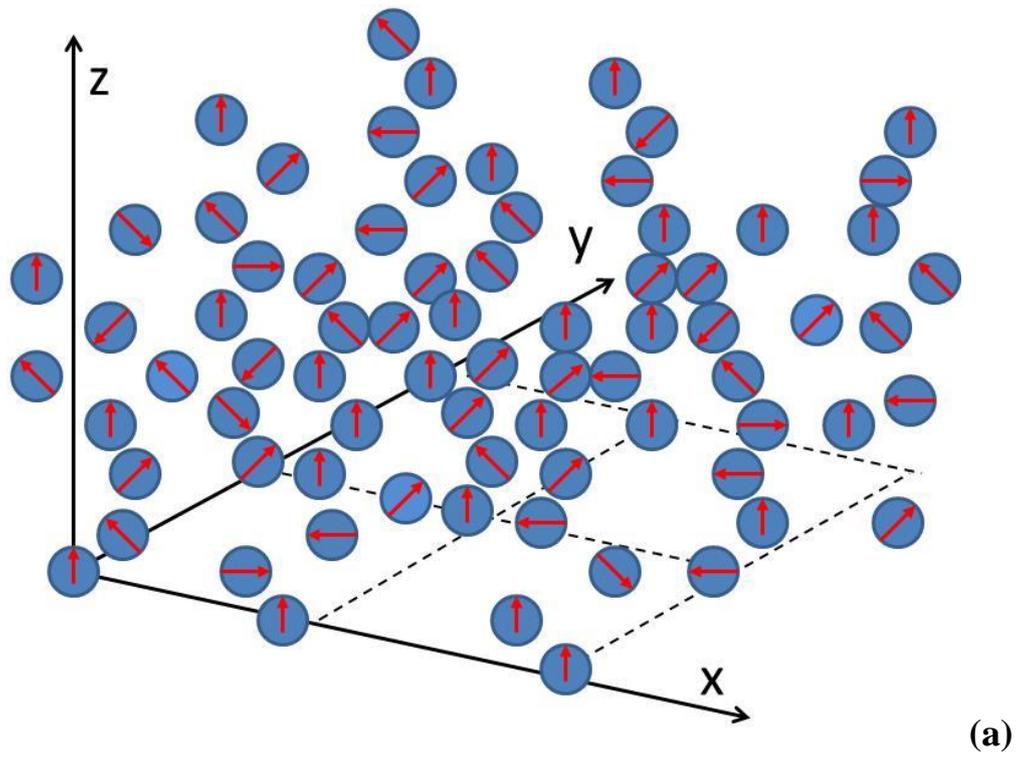

(a)

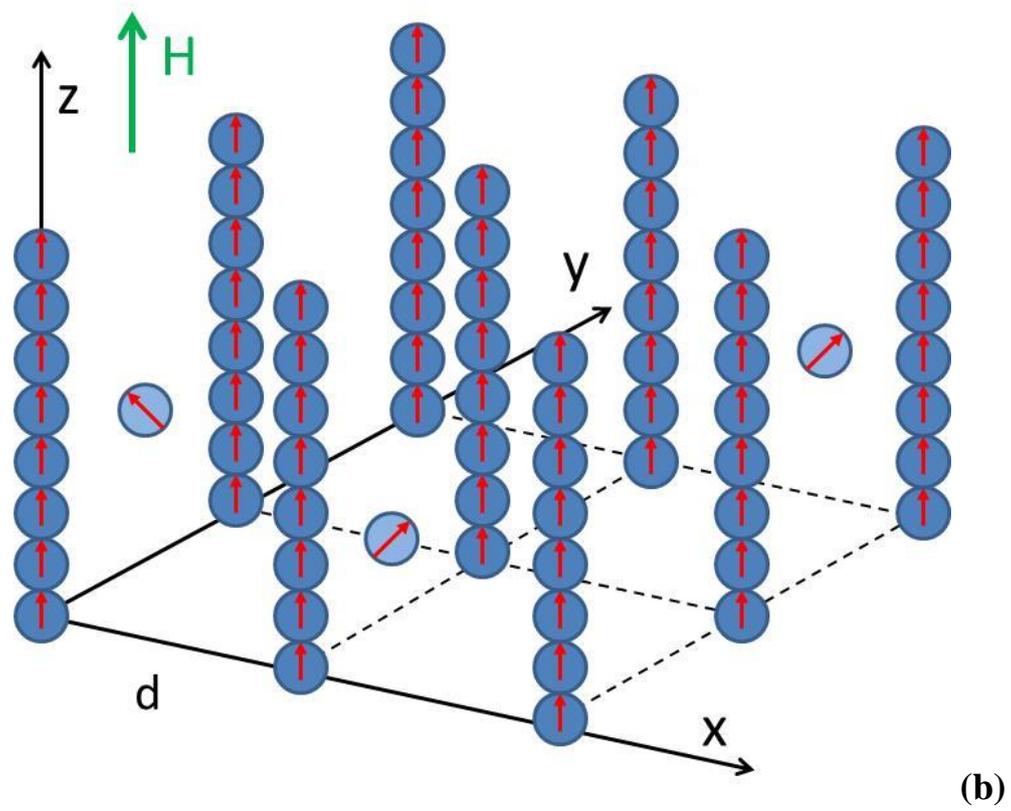

(b)

**Fig. 1**

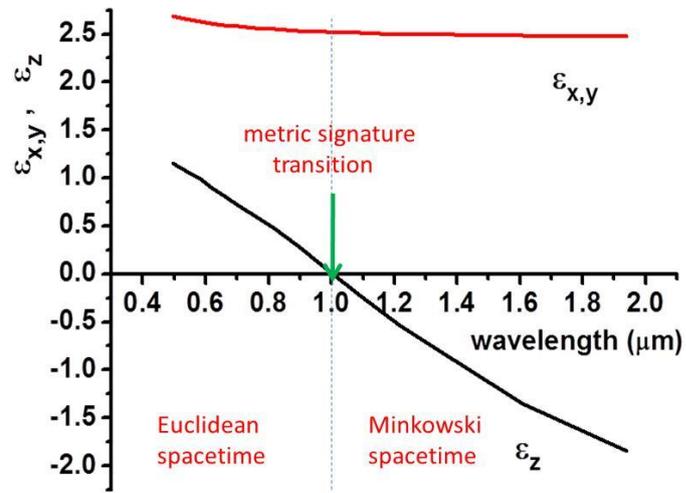

(a)

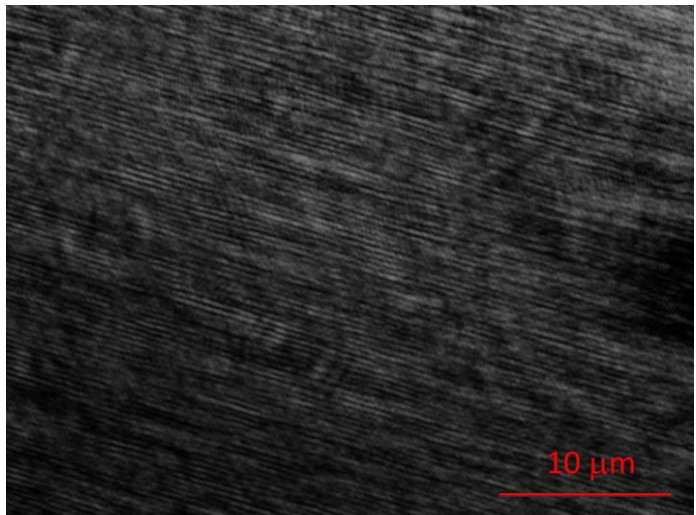

(b)

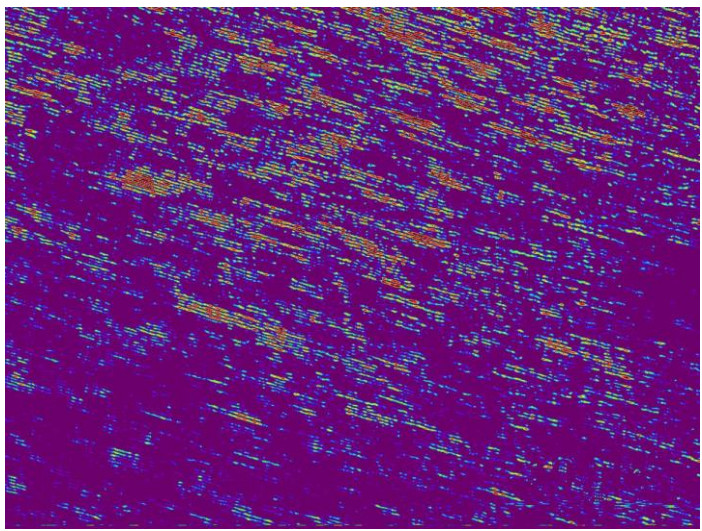

(c)

**Fig. 2**

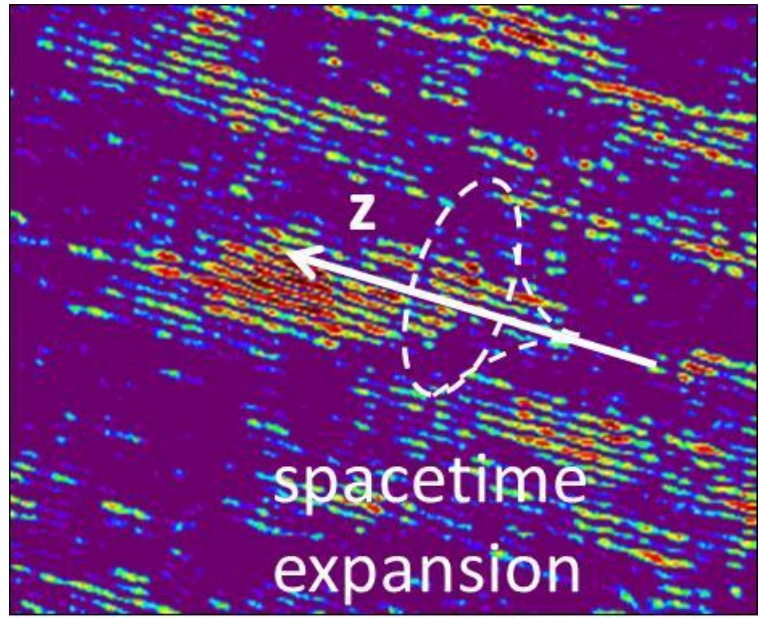

(a)

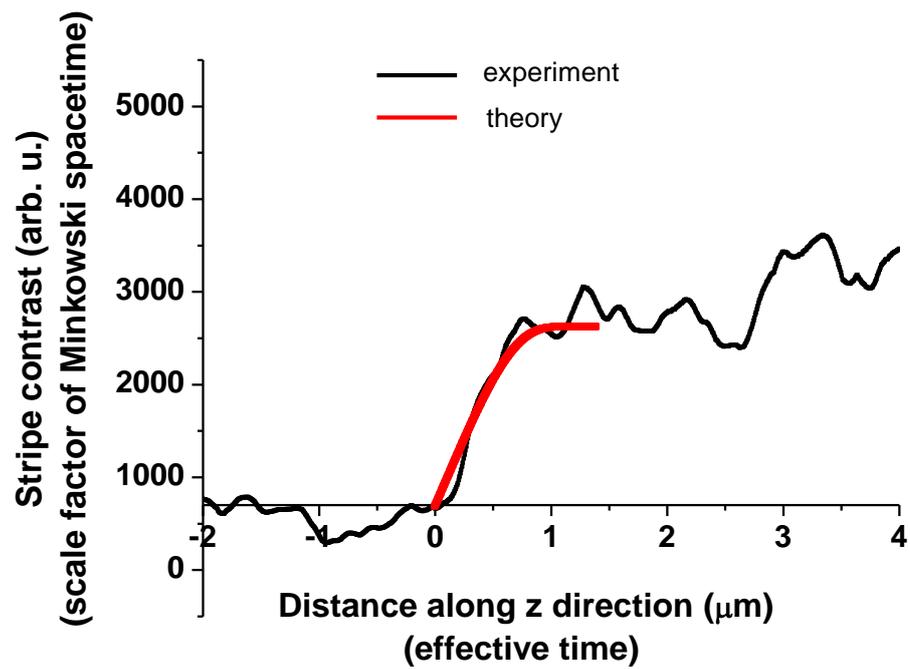

(b)

**Fig. 3**

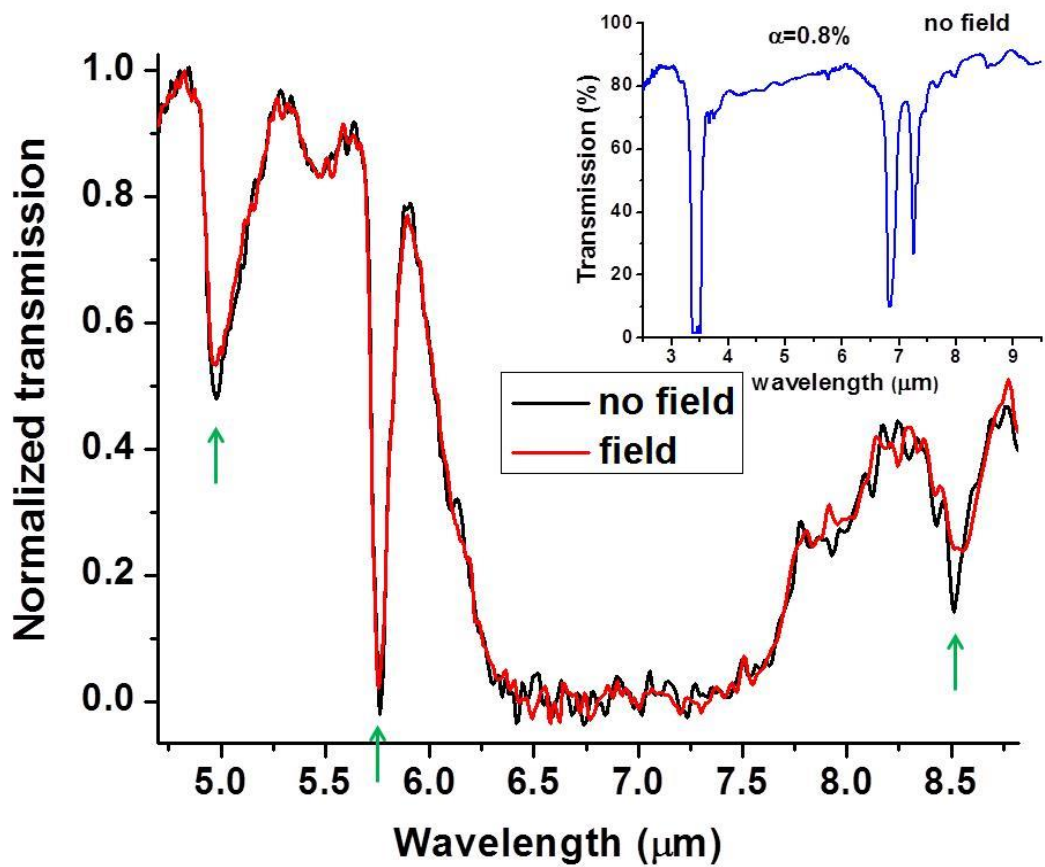

Fig. 4

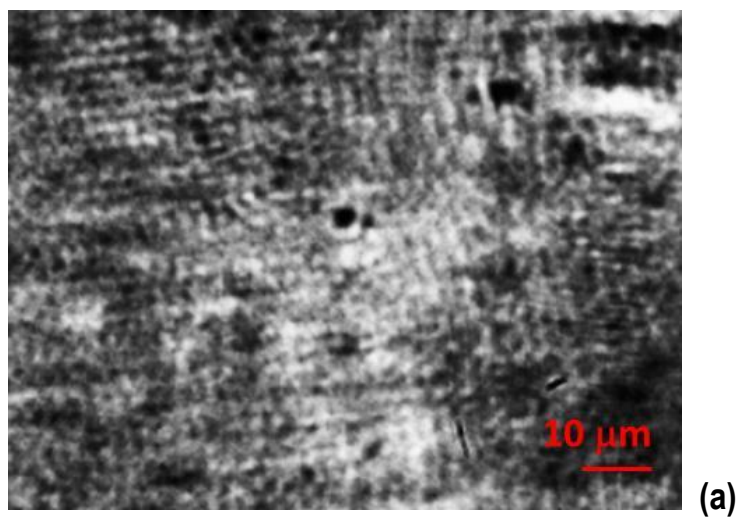

(a)

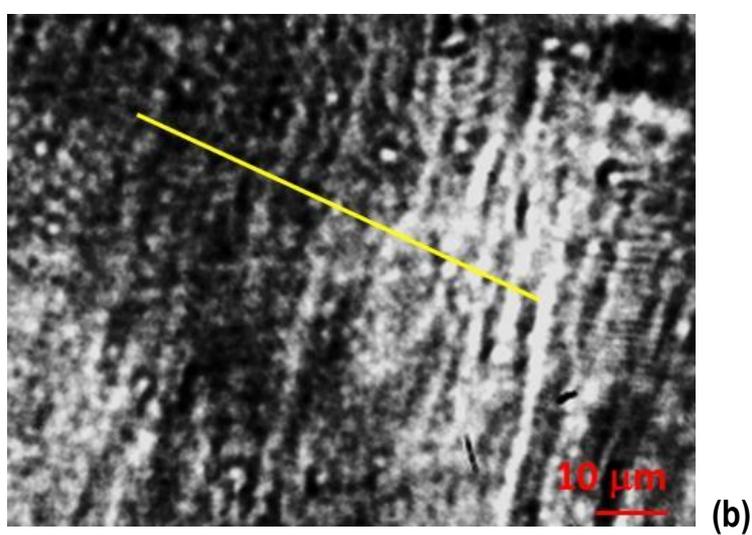

(b)

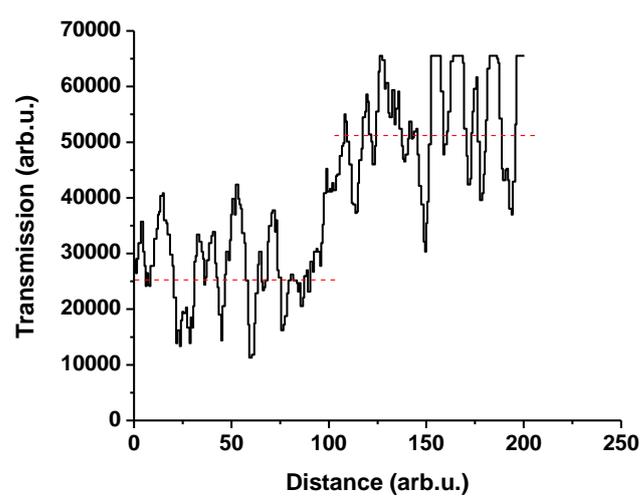

(c)

**Fig. 5**